\documentclass[aps,floats,prd,twocolumn,nofootinbib,preprintnumbers,superscriptaddress,10pt]{revtex4-1}
 
\usepackage{amsfonts,amsmath,amssymb,bm,tensor}
\usepackage{comment}
\usepackage{ifpdf}
\usepackage{slashed}
\usepackage{color}
\usepackage[mathscr]{eucal}
\usepackage[utf8]{inputenc}
 
\usepackage{cancel}

\ifpdf
  \usepackage{graphicx}     
  \usepackage[bookmarksopen,colorlinks=true,linkcolor=bblue,citecolor=bblue,urlcolor=ppink]{hyperref}
\else     
  \usepackage[dvipdfmx]{graphicx}     
  \usepackage[dvipdfmx,bookmarksopen,colorlinks=true,linkcolor=bblue,citecolor=ppink,urlcolor=darkred]{hyperref}
\fi

\definecolor{red}{rgb}{1,0,0}
\definecolor{darkred}{rgb}{0.6,0,0}
\definecolor{darkgreen}{rgb}{0.992447,0.623778,0.034597}
\definecolor{ppink}{rgb}{1,0.4,0.4}
\definecolor{bblue}{rgb}{0.284602,0.317763,0.963947}

\newcommand{\vev}[1]{ \left< {#1} \right> }

\newcommand{\dd}{\mathrm{d}}

\newcommand{\hatn}{{\hat{\bm n}}}

\makeatletter
\newcommand\footnoteref[1]{\protected@xdef\@thefnmark{\ref{#1}}\@footnotemark}
\makeatother

\begin{document}


\title{Chiral photons from chiral gravitational waves}
\author{Keisuke Inomata}
\affiliation{ICRR, University of Tokyo, Kashiwa, 277-8582, Japan}
\affiliation{Kavli IPMU (WPI), UTIAS, University of Tokyo,
Kashiwa, 277-8583, Japan}
\author{Marc Kamionkowski}
\affiliation{Department of Physics and Astronomy, Johns Hopkins University, 3400 N.\ Charles Street,
Baltimore, MD 21218, U.S.A.}

\begin{abstract}
We show that a parity-breaking {\it
uniform} (averaged over all directions on the sky) circular
polarization of amplitude $V_{00} \simeq 2.6 \times 10^{-17}\, \Delta \chi (r/0.06)$ can be induced by chiral gravitational-wave (GW) background with tensor-to-scalar ratio $r$ and chirality parameter $\Delta\chi$ (which is $\pm1$ for a maximally chiral background).  We also show, however, that
a uniform circular polarization can arise from a realization of
a {\it non}-chiral GW background that
spontaneously breaks parity.  The magnitude of this polarization is
drawn from a distribution of root-variance $\sqrt{\vev{ V_{00}^2}} \simeq 1.5\times 10^{-18}\,
(r/0.06)^{1/2}$ implying that the chirality parameter must be $\Delta \chi \gtrsim 0.12 (r/0.06)^{-1/2}$
to establish that the GW background is chiral. 
Although these values are too small to be detected by any
experiment in the foreseeable future, the calculation
is a proof of principle that cosmological
parity breaking in the form of a chiral gravitational-wave
background can be imprinted in the chirality of the photons in
the cosmic microwave background.  It also illustrates how a
seemingly parity-breaking cosmological signal can arise from
parity-conserving physics.
\end{abstract}

\date{\today}
\maketitle
\preprint{IPMU 18-0185}

The cosmic microwave background (CMB) is linearly polarized as a
consequence of the anisotropic Thomson scattering of CMB photons
that arises at linear order in the amplitude of cosmological
perturbations \cite{Rees:1968}.  This Thomson scattering does not, however,
induce circular polarization.  Recent work has shown that
circular polarization can be generated, from primordial
perturbations through post-recombination photon-photon
scattering
\cite{Motie:2011az,Sawyer:2014maa,Ejlli:2016avx,Shakeri:2017iph,Sadegh:2017rnr,Montero-Camacho:2018vgs}, at second order in the density-perturbation
amplitude.  Still, the {\it mean} value of
the circular polarization, when averaged over the sky, is
predicted to be zero.  Since a nonzero circular polarization implies a
particular handedness, the absence of a uniform circular
polarization can be seen as a consequence of the absence of
parity breaking in primordial density perturbations.

Here we show that a uniform circular polarization can arise if
parity is broken in the form of a chiral
primordial gravitational-wave (GW) background.  Chiral GWs
may arise if, for example, there is a Chern-Simons coupling of
the inflaton to gravity \cite{Lue:1998mq}, from gravity at a
Lifshitz point \cite{Takahashi:2009wc}, graviton self-couplings
\cite{Maldacena:2011nz,Anber:2012du}, gaugeflation and
chromo-natural inflation
\cite{Maleknejad:2012fw,Adshead:2013qp,Obata:2016tmo,Bielefeld:2014nza},
Holst gravity \cite{Contaldi:2008yz}, and in models that connect
leptogenesis to primordial gravitational waves
\cite{Alexander:2004us,Abedi:2018top}. 
 The circular polarization arises
through interactions of linearly polarized CMB photons with CMB
anisotropies along the line of sight to the surface of last
scatter.  The correlation between
the photon anisotropy induced by the gravitational wave and the
primordial linear polarization also induced by the gravitational
wave leads, if the GW background is chiral, to a uniform circular
polarization.

We also show, however, that a uniform circular polarization can
arise as a statistical fluctuation---a cosmic variance---if the
stochastic GW background is {\it not} chiral.  Even if the expected
value is zero, there will be some variation, in any given
realization of the GW background, in the amplitudes of the
right- and left-handed gravitational waves.  The theory
predicts that the uniform circular polarization $V_{00}$ (where
the $00$ indices indicate the $l=0,m=0$ spherical-harmonic
coefficient of the circular-polarization pattern) is selected
from a distribution centered on $V_{00}=0$ but with nonzero
variance.  
 As we show, though, such a nonzero value for
the uniform circular polarization cannot arise from primordial
density perturbations, and so a nonzero uniform value would
indicate a GW background, even if not necessarily chiral.

Although the circular polarization induced by GWs has been discussed in terms of the photon-graviton scattering~\cite{Bartolo:2018igk}, we focus on the circular polarization induced through photon-photon scattering.
This work builds upon a detailed analysis presented in
Ref.~\cite{Inomata:2018vbu} of the circular polarization induced
by GWs.  That work builds upon a re-calculation
\cite{Kamionkowski:2018syl}, obtained with the TAM formalism
\cite{Dai:2012bc,Dai:2012ma}, of circular polarization induced
by photon-photon scattering, in
Ref.~\cite{Montero-Camacho:2018vgs}, which itself extends
several earlier analyses \cite{Motie:2011az,
Sawyer:2014maa,Ejlli:2016avx,Sadegh:2017rnr} of this effect.
Note that the circular polarization induced through the photon-photon scattering is much larger than that induced through the photon-graviton scattering~\cite{Inomata:2018vbu}.

As discussed in that earlier work, circular polarization is
induced through Faraday conversion as a linearly polarized light
ray propagates through a medium with an anisotropic
index-of-refraction tensor.  The circular
polarization $V(\hatn)$ in direction $\hatn$ is
\begin{align}
     V(\hat{\bm n}) = \epsilon_{ac} P^{ab} (\hat{\bm n})
     \Phi_b^{\ c} (\hat{\bm n}).
\label{eq:v_formula_2}
\end{align}
Here,
\begin{equation}
P_{ab}(\hat{\bm n}) = \frac{1}{\sqrt{2}} \left(
    \begin{array}{cc}
     Q(\hat{\bm n}) & U(\hat{\bm n})  \\
     U(\hat{\bm n}) & -Q(\hat{\bm n}) \\
    \end{array}
  \right),
\end{equation}
is the polarization tensor, whose components are the Stokes
parameters $Q(\hat{\bm n})$ and $U(\hat{\bf n})$, and
\begin{equation}
\Phi_{ab}(\hat{\bm n}) = \frac{1}{\sqrt{2}} \left(
    \begin{array}{cc}
     \phi_Q(\hat{\bm n}) & \phi_U(\hat{\bm n})  \\
     \phi_U(\hat{\bm n}) & -\phi_Q(\hat{\bm n}) \\
    \end{array}
  \right),
\end{equation}
is a phase-shift tensor that describes the phase shifts induced
by the index-of-refraction tensor.  These are obtained as
line-of-sight integrals,
\begin{align}
     \phi_{Q,U}(\hat{\bm n}) = \frac{2}{c}
     \int^{\chi_\text{LSS}}_0 \frac{\dd \chi}{1+z} \,
     \omega(\chi) n_{Q,U}(\hat{\bm n} \chi),
\label{eq:phi_qu_def}
\end{align}
where $z$ is redshift and $\chi_\text{LSS}$ the comoving
distance to the last-scattering surface.  Here, $n_Q$ and $n_U$
are components, in a plane transverse to the line of sight, of
the index-of-refraction tensor,
\begin{equation}
n_{ab} = \delta_{ab} + \frac{1}{2} ( \chi_{e,ab} + \chi_{m,ab} ) = \left(
    \begin{array}{cc}
     n_I + n_Q & n_U + in_V  \\
     n_U - in_V & n_I - n_Q \\
    \end{array}
  \right).
\end{equation}
Finally, $\epsilon_{ab}$ is the antisymmetric tensor on the
2-sphere.

The polarization and phase-shift tensors can both be expanded,
\begin{align}
\label{eq:p_tensor}
     P_{ab}(\hat{\bm n}) &= \sum_{lm} \left[P^E_{lm}
     Y^{TE}_{(lm)ab}(\hat{\bm n}) + P^B_{lm}
     Y^{TB}_{(lm)ab}(\hat{\bm n})\right], \\
\label{eq:phi_tensor}
     \Phi_{ab}(\hat{\bm n}) &= \sum_{lm} \left[\Phi^E_{lm}
     Y^{TE}_{(lm)ab}(\hat{\bm n}) + \Phi^B_{lm}
     Y^{TB}_{(lm)ab}(\hat{\bm n}) \right],
\end{align}
in terms of tensor spherical harmonics $Y^{TE}_{(lm)ab}(\hat{\bm n})$ and $Y^{TB}_{(lm)ab}(\hat{\bm
n})$~\cite{Kamionkowski:1996ks,Zaldarriaga:1996xe,Kamionkowski:2015yta}.
The expansion coefficients
in Eqs.~(\ref{eq:p_tensor}) and (\ref{eq:phi_tensor}) are random
variables chosen from a distribution of zero mean and variances
$C_l^{XX} = \vev{ \left|X_{lm}\right|^2}$ for
$X=P^E,P^B,\Phi^E,\Phi^B$.  In the absence of parity breaking,
there are, moreover, cross-correlations $C_l^{P^E \Phi^E}$ and
$C_l^{P^B \Phi^B}$.  If there is no parity breaking, then there
is no cross-correlation between the E modes and the B
modes---the two have opposite parity and any cross-correlation
would imply a preferred handedness.  Expressions for all the
power spectra $C_l^{XY}$ are given as integrals over the
primordial gravitational-wave power spectra in
Ref.~\cite{Inomata:2018vbu}.

The coefficients $V_{lm} = \int \dd \hat{\bm n} V(\hat{\bm n}) Y^*_{lm}(\hat{\bm n})$ in the spherical-harmonic expansion of the circular polarization can be expressed in terms of $P_{lm}$ and $\Phi_{lm}$ as~\cite{Inomata:2018vbu}
\begin{align}
V_{lm} =& \sum_{l_1 m_1 l_2 m_2 (\rm{odd})}
\left( P_{l_1 m_1}^E \Phi_{l_2 m_2}^E ( i G^{l\, -m}_{l_1 m_1 l_2 m_2} )  \right. \nonumber \\
& \qquad \qquad \qquad \qquad 
\left. + P_{l_1 m_1}^B \Phi_{l_2 m_2}^B ( i G^{l\, -m}_{l_1 m_1 l_2 m_2} ) \right) \nonumber \\
&+  \sum_{l_1 m_1 l_2 m_2 (\rm{even})} 
\left( P_{l_1 m_1}^E \Phi_{l_2 m_2}^B ( -G^{l\, -m}_{l_1 m_1 l_2 m_2} ) \right. \nonumber \\
& \qquad \qquad \qquad \qquad 
\left. - P_{l_1 m_1}^B \Phi_{l_2 m_2}^E ( -G^{l\, -m}_{l_1 m_1
l_2 m_2} ) \right),
\label{eqn:vlm}
\end{align}
where the subscripts (odd) and (even) indicate summations over $l_1 + l_2 + l = \text{odd}$ and $l_1 + l_2 + l = \text{even}$, respectively.
Here $G^{lm}_{l_1 m_1 l_2 m_2} = - \xi^{lm}_{l_1\, -m_1,
l_2 m_2} H^l_{l_1 l_2}$, where $\xi^{lm}_{l_1\, m_1,
l_2 m_2}$ and $H^l_{l_1 l_2}$ are defined in terms of Wigner 3-j symbols as
\begin{align}
 \xi^{lm}_{l_1m_1 l_2m_2} &\equiv (-1)^m \sqrt{ \frac{(2l_1+1) ( 2l +1) (2l_2 +1)}{4\pi}} 
 \left( 
 \begin{array}{ccc}
      l_1 & l & l_2  \\
      -m_1 & m & m_2
    \end{array}
  \right), \\
  H^l_{l_1 l_2} &\equiv  \left( 
 \begin{array}{ccc}
      l_1 & l & l_2  \\
      2 & 0 & -2
    \end{array}
  \right).
\end{align}
In Ref.~\cite{Inomata:2018vbu} we considered anisotropies in the circular polarization and found that the circular polarizations induced by density (scalar metric) perturbations are much larger than those induced by tensor perturbations.
However, scalar perturbations induce no \emph{uniform} circular polarization $V_{00}$.
We therefore focus here on $V_{00}$, as it provides a clean signature of tensor perturbations.
Setting $l=m=0$ and taking $G^{00}_{l_1m_1 l_2 m_2} = \delta_{l_1l_2}\delta_{-m_1m_2}/\sqrt{4\pi}$, we express the uniform circular polarization as
\begin{equation}
     V_{00} = \frac{1}{\sqrt{4\pi}} \sum_{lm} \left( P_{lm}^E \Phi_{lm}^{B\, *} - P_{lm}^B
     \Phi_{lm}^{E\,*}  \right).
\label{eqn:uniformrotation}     
\end{equation}
Taking the expectation value, over all realizations of metric
perturbations, we find,
\begin{equation}
     \vev{V_{00}} = \frac{1}{\sqrt{4\pi}} \sum_{lm} \left( C_l^{P^E
     \Phi^B} - C_l^{P^B
     \Phi^E} \right).
\end{equation}
Through calculations that parallel those in
Ref.~\cite{Inomata:2018vbu} and lead to the power spectrum
$C_l^{P^E \Phi^B}$ induced by photon-photon scattering,
we infer power spectra,
\begin{align}
     C_l^{P^E \Phi^B} &= 4\pi A \int \frac{\dd k}{k} 2  \left(\frac{k^3}{2\pi^2}P^{(TE,TB)} (k) \right)  \nonumber\\
    & \times \int_0^{\eta_0}\, \dd\eta\, g(\eta) \left(
     -\sqrt{6} \mathcal{P}^{(2)}(k,\eta) \right) \epsilon_l^{(2)} ( k
     (\eta_0-\eta)) \nonumber \\
    & \times \int^{\eta_0}_{\eta_{\text{lss}}}\, \dd\eta
     \, (1+z)^4 \left( \bar a_{2,2}^E(k,\eta) \right)
     \beta_l^{(2)} ( k (\eta_0-\eta)).
\label{eq:c_pephib}     
\end{align}
The coefficient $A$ is inferred from the Euler-Heisenberg Lagrangian and provided in Ref.~\cite{Montero-Camacho:2018vgs}. 
The second and third lines in Eq.~(\ref{eq:c_pephib}) represent the transfer functions of $P^E$ and $\Phi^B$ respectively.
In particular, $g(\eta)$ is the visibility function, $\mathcal P^{(2)}$ is the function defined in Ref.~\cite{Tram:2013ima}, $\bar a_{2,2}^E$ is the transfer function of the local E-mode moment induced by primordial perturbations, and $\epsilon_l^{(2)}$ and $\beta_l^{(2)}$ are the radial functions coming from the nature of the E-mode and B-mode respectively (see Ref.~\cite{Hu:1997hp} for detail).
Here, $P^{(TE,TB)}(k)$ is defined as the power spectrum for the cross-correlation between the amplitudes $h_{(lm)}^{k,TE}$ and $h_{(lm)}^{k,TB}$,
\begin{equation}
\vev{ h_{(lm)}^{k,TE} \left(h_{(lm)}^{k',TB} \right)^* } = i \frac{(2\pi)^3}{k^2} \delta(k-k') P^{(TE,TB)},
\label{eq:h_tam_eb}
\end{equation}
where $h_{(lm)}^{k,TE}$ and $h_{(lm)}^{k,TB}$ are the TE and TB modes in the TAM
decomposition for the transverse-traceless metric perturbation.
The $i$ in Eq.~(\ref{eq:h_tam_eb}) is canceled out by the $i$ in front of $\beta^{(2)}_l$ in Eq.~(41) in Ref.~\cite{Inomata:2018vbu}.
$C_l^{P^B \Phi^E}$ are also given by the same equation except for $\epsilon_l^{(2)} \leftrightarrow \beta_l^{(2)}$.
The TE and TB modes have opposite parity, and so any nonzero correlation between them indicates parity breaking.
Such a cross-correlation arises if parity is broken by a disparity in the amplitudes of the right- and left-circularly polarized gravitational waves.
This can be seen by writing the amplitudes,
\begin{equation}
     h_{(lm)}^{k,\pm} = \frac{1}{\sqrt{2}} \left(h_{(lm)}^{k,TE} \mp
     i\, h_{(lm)}^{k,TB} \right),
\label{eqn:helicitybasis} 
\end{equation}
for the $\pm$ (R and L) helicity-basis TAM
waves (cf., Eq.~(14) in Ref.~\cite{Inomata:2018vbu}).  
Following Ref.~\cite{Gluscevic:2010vv}, we define the chirality parameter
$\Delta\chi$ through $P_\pm(k) = (1\mp\Delta\chi)P_T(k)$, where
$P_T(k)$ is the primordial GW power spectrum, and 
$P_+$ and $P_-$ correspond to $P_\text{R}$ and $P_\text{L}$ in Ref~\cite{Gluscevic:2010vv}, respectively.
 From this and Eq.~(\ref{eqn:helicitybasis}), it follows that 
\begin{equation}
     {P}^{(TE,TB)}(k) = \Delta \chi P_T(k).
\end{equation}

As a result, we find by numerical evaluation the uniform
circular polarization, normalized by the physical units $T_\text{CMB} = 2.7255$\,K~\cite{Fixsen:2009ug}, to be
\begin{align}
     \vev{V_{00}} &= \sum_{l} \frac{2l+1}{\sqrt{4\pi}} (C_l^{P^E
     \Phi^B} - C_l^{P^B
     \Phi^E}) \nonumber \\
     &\simeq  2.6\times 10^{-17} \, {\Delta \chi}
     \left(\frac{r}{0.06} \right),
\label{eqn:PBnumerical}     
\end{align}
for a gravitational-wave background with tensor-to-scalar ratio
$r$ ($\lesssim 0.06$ \cite{Aghanim:2018eyx, Ade:2018gkx}) and chirality parameter $\Delta\chi$.
This numerical result is obtained assuming a scale-invariant power spectrum and chirality. Although the analysis of a non-scale-invariant spectrum is beyond the scope of this paper, we do not expect any qualitative difference for non-scale-invariant spectra consistent, at the relevant length scales, with observational constraints.

So far, we have calculated the {\it expectation value} for $V_{00}$.
However, the {\it observed} value of the circular polarization
arises as the result of a single realization of a random field.
The prediction is thus that $V_{00}$ is selected from a random
distribution, with some nonzero variance $\vev{(\Delta V_{00})^2} =
\vev{V_{00}^2}-\vev{V_{00}}^2$, with an expectation value $\vev{V_{00}}$. 
Note that, in the absence of the parity breaking, $\vev{(\Delta V_{00})^2} = \vev{V_{00}^2}$ is satisfied.
From Eq.~(\ref{eqn:uniformrotation}), we find,
\begin{align}
    \label{eq:v_cos_vari_formula}
     \vev{(\Delta V_{00})^2} &=\sum_l \frac{2l + 1}{4 \pi} \left( C_l^{P^E P^E}
     C_{l}^{\Phi^B \Phi^B} + C_{l}^{P^B P^B} C_{l}^{\Phi^E
     \Phi^E} \right. \nonumber \\
     &  - 2 C_{l}^{P^E \Phi^E} C_{l}^{P^B \Phi^B}
      \nonumber \\
     & \left. +\left(C_l^{P^E \Phi^B} \right)^2 +\left(C_l^{P^B
     \Phi^E} \right)^2 - 2 C_l^{P^E P^B} C_l^{\Phi^E\Phi^B}
     \right).
\end{align}
There are contributions---those in the first and second lines---that
arise even in the absence of parity breaking (while those that
arise from parity-breaking physics are listed in the third
line).  Thus, the observed Universe can have a
uniform circular polarization as a consequence of a 
parity-breaking realization of scalar and tensor metric 
perturbations, even in the absence of parity breaking in the
underlying physics (in analogy to spontaneous symmetry breaking
in particle theory).  Such a
possibility does not arise if there are only density
perturbations (and thus only E-mode polarization and
index-of-refraction tensor); a uniform circular polarization
requires tensor perturbations (and thus B modes).
As a result, we find numerically,
\begin{equation}
     \vev{(\Delta V_{00})^2}^{1/2}  \simeq 1.5 \times 10^{-18}\, \left(
     \frac{r}{0.06} \right)^{1/2}.
     \label{eq:v_cos_vari}
\end{equation}
Comparing Eq.~(\ref{eq:v_cos_vari}) with Eq.~(\ref{eqn:PBnumerical}), we see that the chirality parameter must be  $\Delta \chi \gtrsim 0.12  (r/0.06)^{-1/2}$ if detection of a uniform circular polarization can be attributed, at the $2\sigma$ level, to a chiral GW background.  Although detection of a nonzero uniform circular polarization with $V_{00} \lesssim 3\times 10^{-18}\, (r/0.06)^{1/2}$ would not necessarily indicate a chiral GW background, it would still indicate the presence of tensor (or perhaps vector) perturbations.
It is interesting to understand why the
uniform circular polarization $\vev{V_{00}}$ for
maximal chirality ($\Delta \chi = 1$) is roughly 20 times the
root-variance $\vev{(\Delta V_{00})^2}^{1/2}$ in the absence of
any chirality.  To do so, we first note that the numerical
result in Eq.~(\ref{eqn:PBnumerical}) arises from Faraday
conversion by GWs of a primordial linear polarization
that is also induced by GWs, while the variance
in Eq.~(\ref{eq:v_cos_vari}) arises primarily from Faraday
conversion by GWs of primordial linear
polarization induced by density perturbations (and {\it vice
versa}).  The contribution to the variance from Faraday
conversion of GW-induced linear polarization by
GWs turns out to be $\vev{(\Delta
V_{00})^2}^{1/2} \simeq 1.5\times 10^{-19}(r/0.06)$ (smaller by
$\sim 10$ given that the relevant primordial linear polarization
induced by GWs is $\sim 0.1$ times that
induced by density perturbations for $r\sim0.1$).  
If $\sim N_{\rm gw}$ GW modes are contributing to the
signal, then we expect, on average, $N_{\rm gw}/2$ to be right
handed and a similar number left handed.  
Still, there will be root-$N$ fluctuations in both signals from right-handed and left-handed GWs in any given realization
of a non-chiral GW background, implying a variance $\sim N_{\rm
gw}^{-1/2}$ times the expectation value in the maximally-chiral
case. 
From the numbers reported above, this suggests $N_{\rm
gw}\sim 10^4$, or that the uniform circular polarization is
(given that there are $2l+1$ modes for each $l$) dominated by GW
modes with multipole moments $l\lesssim 100$.  We have verified
numerically that this is the case.

In practice, these values of the circular polarization are too
small to be detected in the foreseeable future.  Still, the result
is a proof of principle that gravitational-wave chirality can be
imprinted in the chirality of the cosmic microwave background.
Measurement of the circular
polarization would, in the event of detection of a non-zero EB
correlation in the CMB polarization, help distinguish a
chiral-GW explanation~\cite{Saito:2007kt} for such an effect~\cite{Lue:1998mq,Gluscevic:2010vv,Ferte:2014gja}
from cosmic birefringence.  It would also complement probes of
the GW chirality at nanoHertz frequencies \cite{Qin:2018yhy} and
at LIGO/LISA frequencies \cite{Seto:2007tn,Smith:2016jqs}.


We thank Daniel Green for useful discussion.
K.I. is supported by World Premier International Research Center
Initiative (WPI Initiative), MEXT, Japan, Advanced Leading
Graduate Course for Photon Science, and JSPS Research
Fellowship for Young Scientists, and thanks Johns Hopkins
University for hospitality.  This work was supported at Johns
Hopkins by NASA Grant No.\ NNX17AK38G, NSF Grant
No.\ 1818899, and the Simons Foundation.


\end{document}